\documentstyle[12pt,russian]{article}
\begin{document}
\centerline {\bf  q-Plastic deformation}
\centerline {\it Trinh Van Khoa}
\centerline {\it Department of physics of polymers and crystal,
MGU}
\centerline {\it E-mail : khoa@polly.phys.msu.ru }
\bigskip
\large

1. Let's consider a continuum with the structure of knots (for
example, it is the polymers [18]). It is known that under an
action of a force, the media transferees to  plastic state. All
of the experiments about plasticity had showed that in the plastic
state the fields of stress and strains was in strongly
fluctuation. Furthermore, the diagram $\gamma=f(e)$ of the
stress-strain is similar to the diagram $P=f(V)$ of the
pressure-volume in the fluid [1]. In the other side one observes
the phase transition in the neighborhood of the end of the crack
[2]. So we may consider the process of plastic deformation as
critical one in which the formation of the structure was appeared
[6]. The density of probability for the transition to the
plasticity in the solution of the equation Fokker-Planck. However
with the help of potential character of the deformation process we
may use Schrodingers equation in the role of Fokker-Planck one. It
mean that we may consider the plastic process as one with
supersymmetry. Where we take the deformation tensor in the role of
order parameter. In paper [17] A.A. Iliushin proposed the
conception of "trajectory of deformation "in the space $E_5$ (it is
the image of the trajectory of plastic wave or of the trajectory
of deformation in usual sense). As a matter of fact the space
$E_5$ is the one fiber in the deformation bundle [6].
 In other
word the deformation field is gauge one, in which generalized
trajectory Iliushin was their  section.
Space of deformation $P = \{E_n(x, t) \} $, with most general
 points of view, is noncommutative.
In the paper [6] we constructed the noncommutative model of
deformation. In frameworks of this model we give a proof of the
Iliusin hypothesis and Koiter hypothesis of existence of
singularities on the limiting surface of deformation. Iliusin
hypothesis - it is the invariant of the simplectic structure under
the diffeomorfizm . The existence of Koiter singularities has a
topological reason and number of singularities it is topological
number Pontriagin. It is known, that the space of deformation $P =
\{E-n \} $ is associative algebra,
 in frameworks of which  can be determined derivative and
the differential form.
So, we choose universal algebra
Connes $ \Omega (P) $ as algebra of the differential forms above
 space of deformation $P $. Thus, differential
 calculation on $ \Omega_D (P) $ has a type similar to such in work
[9], according to which least differential algebra of algebra of
cosicle $C(Der(P),P) $, containing the space $P $, is $
\Omega_D(P) $, being algebra $C (Der(P),P) $, where $Der(P) $ -
algebra of derivative on $P $. In this paper we will construct the
model of deformation, in which universal algebra Connes is algebra
Hoof. The differential calculus on this algebra has a type of
Woronowicz.
 When we identify the
trajectory of deformation (or the line of propagation of plastic
wave) with the knot we have archived the density of probability of
the to plasticity [6]:
$$
Z=\int exp[-kS_{cs}]de,\eqno(*)
$$
where $k-$ it is elastic coefficient, $S_{cs}$- the action
Chern-Sinomce $S_{cs}=\int\Gamma_{cs}$, and $\Gamma_{cs}$- the
form Chern-Simonce for deformation variety.

2. In the study of the exact solved system a new algebraic
structure was appeared. It was called by the quantum group. Though
quantum group in exact mathematical sense b ut there are many of
its characters that one may to consider as a group. That's why
there is a problem to study and to construct the physical and
mechanical models in which quantum group is a symmetry one. As a
matter of fact, noncommutative geometry and quantum group are most
effective  instruments to study the deformation process.
Corresponding to every process of deformation $E(x,t)$ there is a
deformation space $P=\{E(x,t)\}$. In this process there is a
reforming of symmetry and a generation of order structure. The
coherence state in the block Bobrokov - Revuzenko - Shemiakin was
their experimental affirm [6]. In most general view it is the
noncommutative space. Furthermore, may be, it is a q-space of
deformation. So in the deformation process there are
noncommutative bundle and the fluctuation.
Now our problem - it is determining functional $Z$ in (*).
In the classical case
[7] "secondary calculus" was defined by formalism BRST, in which
the space of state $P$ was extended to the space
$\tilde{P}=(x,y,\bar{y})$, where $x$- spacetime coordinate,
$y,\bar{y}$- noncommutative coordinate. In this space one defined
1-form $\tilde{A^a}$ with the value in the algebra Lie $\cal G$ of
gauge group $G$: $\tilde{P}=(x,y,\bar{y})= A^a_{mu}dx^{mu}+
A^a_{y}dy + A^a_{\bar{y}d\bar{y}}$. So $A^a_{\mu}(x,o,o)$ was
identified with usual gauge field, the ghost and antighost are
fields $C^a(x)= A^a_{y}(x,o,o)dy$,
$\bar{C^a(x)}=A^a_{\bar{y}}(x,o,o)d\bar{y}$. Thus one may have
been generalized tensor gauge field by the 2-form
$\tilde{B^a}(x,y,\bar{y})=\frac{1}{2}B_{\mu\nu}dx^{mu}\wedge
dx^{nu} + B_{\mu y}dx^{\mu}\wedge dy + \frac{1}{2}B_{yy}dy\wedge
dy + B_{y\bar{y}}dy\wedge d\bar{y} +
\frac{1}{2}B_{\bar{y}\bar{y}}d\bar{y}\wedge \bar{y}$. However in
the noncommutative space of deformation it is necessary for me to
realize only the first stage of this procedure because the gauge
1-form on $P$ was the differential form with value in matrix
algebra of deformation. Thus one can formulate q-BRST cohomology
of plastic deformation in the formalism Watamura [8]. It is the
comodule algebra upon $Fund_q(G)$ that was generalized from
component comodule $A=C(C^I,A^I,DA^I)/{\cal F }$, where $C^I$- is
a ghost, $A^I-$ a gauge field, $\cal F-$ the commutative covariant
correlations. Of course, because of the expansion
$P\rightarrow\tilde{A}$ we have to carry out expanding
$d\rightarrow\tilde{d}=d+\delta$, where $d-$ differential operator
on the spacetime and $\delta$- the BRST operator on the field:
$$\delta^2=0,~~d^2=0,~~d\delta+\delta d=0 $$
$$d(XY)=(dX)Y+(-1)^{n_x}X(dY) $$
$$\delta(XY)=(\delta)Y+(-1)^{n_x}X(\delta Y) $$
where $n_x$- the index of the degree of the form and the numbers
of ghost, $X,Y$- are fields. Basic difference between q-BRST
deformation and classical BRST [7] is the effect that the
differential calculus on the expanded space of deformation
$\tilde{A}$ is bicovariant. In this case we have been used
differential calculus in [9] for the group $SU_q(N)$ and
$SO_q(N)$. It is the differential calculus of the Woronowicz [4]
in the limits of the noncommutative geometry Connes [3].Thus first
of all, the bicovariant bimodule $\Gamma$ on the expanded space of
deformation $\tilde{A}$ was built. It is left and right covariant
$A-$ modul. Letter fundamental bimodul was given. It is the
bicovariant bimodul that is linear subspace with right-covariant
basic $\eta^i$ and coaction $\Delta_L$: $\Delta_L(\eta^i)=
M^i_j\otimes\eta^i$, where $M^i_j$ are matrix of deformation in
space $\tilde{A}$. At last one defined the bicovariant bimodul
$\Gamma_{Ad}$, right-invariant basic of which was adjoint
representation $\theta^i_j=\eta^i_{+}\bar{\eta_{+j}},
~~\bar{\eta}\equiv(\eta^i_{\mp})^{+}$, where $\eta^i_{\mp}$ are
right-invariant basic with the coefficient $f^i_{\pm j}$[9]. As in
the [7] we establish q-BRST equation for the ghost with requiring
that BRST-map of the ghost has to satisfy "horizontal condition",
e.i. it is the equation Cartan-Mayere with covariant differential
calculus. Considering the ghost as right-invariant basic and after
separating simple and adjoint representation for the group $SU(2)$
we obtain
$$
\delta C^a=\frac{-iq}{q^2+q^{-2}}f^a_{ab}C^bC^c,~~\delta C^0=0,
\eqno(1)
$$
Operator $\delta$ acts on the fields of deformation by the rule
[8]
$$ \delta A^I=da^I - ig[C^I,A^I]$$
or
$$
\delta A^a=dC^a-ig(\omega C^0A^0+f^a_{bc}C^bA^c),~~\delta
A^0\eqno(2)
$$
As in classical case the strength of deformation field has form
$$
F^a=dA^0-\frac{-iq}{q^2+q^{-2}}f^a_{bc}A^bA^c,~~F^0=dA^0\eqno(3)
$$
and the equality Bianchi is in the form
$$
dF^a=\frac{-iq}{q^2+q^{-2}}f^a_{bc}[A^bF^c-F^bA^c],~~dF^0=0\eqno(4)
$$
The act of operator $\delta$ was defined as
$$
\delta F^I=\frac{ig}{\omega}[C^0,F^I]
$$
Using this correlation we can build q-class Chern that is similar
to [10,11]
$$
Q=F^IF^J g_{IJ}\eqno(5)
$$
where the form Killing $g_{IJ}$ was defined from the condition
$$
\delta Q=0,~~dQ=0\eqno(6)
$$
Using the permutation operator $\sigma$ and structure constant
$\Lambda$ in [9] and correlation (2),(3) for $F$, we obtain the
other condition (6)for the form Killing
$$
\delta^0_I g_{JK}=\Lambda^{SO}_{TK}\Lambda^{RT}_{IJ}g_{RS},
$$
$$
g_{00}=-g^2\lambda^2[2],~~q^{-1}g_{-+}=qg_{-+}=g_{33}=-q^2(\lambda+2)
$$
Thus q-class Chern is defined in the form
$$
Q(A)=kP, \eqno(7)
$$
where $k$ is q-homotopy operator [11,12]
$$
k=\int^1_0 l_t
$$
and operator $l_t$ is defined by rue:
$$
l_tA^J_t=0~~l_tF^J_t=\hat{\delta_q}A^J_t=d_q
tA^J,~~\hat{\delta_q}\equiv d_q t\frac{\partial}{\partial_q t}
$$
$$
l_t\{f(t)g(t)\}=\{l_tf(t)g(t)\}+(-1)^{n_x}f(q^{-1 t})\{l_t g(t)\}
$$
Parameter $t$ is defined as
$$
A^J_t=tA^J,~F^J_t=tF^J+\frac{ig(t^2-1)}{\lambda}(A^0A^J+A^JA^0),\eqno(8)
$$
After placing (8) into (7) and after carrying out integral refer
$t$ we obtain
$$
Q(A)=\frac{t\lambda}{[2]}<A,dA>+\frac{ig(\lambda^2+2)}{[3]\lambda}
<A,A^0A^J+A^JA^0>
$$

3. The main idea in my approach is to identify the trajectory of
deformation with a knot. Further, similar to the classical
mechanics of continuous medium, using hypothesis in [13] we have
to require in order that strength of deformation field along the
trajectory of deformation to be maximum. As was showed by Witten
in [14] this requirement may be realized when action of
deformation field was the action Chern - Simonce. This action was
defined by $Q(A)$.

Thus we obtain the statistical sum describing the probability of
the transition elastic-plastic:
$$
Z=N\int[-\beta Q_{cs}(A)]dA\eqno(9)
$$
where $\beta$- it is elastic coefficient, $N_1$- coefficient
depending on the potential at beginning and the end point of
trajectory of deformation. If in the medium there is the
dislocation line, then generalized matrix Burgers will be
presented in the form
$$
b_{\Gamma}=Nexp[-\int_C A_{(\mu)}d(x^{\mu}\otimes 1)]\eqno(10)
$$
where $N$- it is order exponent part. In this situation we obtain
statistical sum reflecting the density of probability of the
transition elastic-plastic with the interaction of dislocation
$$
Z=\int\Phi_c(A)exp[-\beta Q_{cs}]dA\eqno(11)
$$
$$
\Phi_c=Tr(b_{\Gamma})
$$

However, it is known that when the medium has been passed on the
plastic state the field of strain and stress will be in the
fluctuation .As a result the action $Q$ will be separated on three
parts:
$$
Q\rightarrow Q_K+Q_{GF}+Q_{FP}
$$
where $Q_K$- classical action Chern-Simonce, $Q_{GF}$- The action
of the fixed gauge and $Q_{FP}$- the action of ghost Fadeev-Popov.
In the formalism of BRST of Watamura [8] this action has to be the
form in order that when $q \to 1$ ones passes to the classical
form in [15]. Then we obtain :
$$
Q_{GF}=\frac{\beta}{4\pi}<dA,dA>,~~Q_{FP}=-dC(DC)
$$
where , similar to [16], $D=d+i[A,A]$. Putting this action into
(9) and (11) we obtain the probability of the transition
elastic-plastic.

4. In this paper we used the formalism BRST to carry out the
procedure of "secondary calculus". But,  the must
interesting for me is the mechanism "quantization" of quantum group . It
is known that the appearance of the order state in continuous
medium is the result of the spontaneous breaking of the symmetry.
The collective fashion [6] was used in the role of  carrier
maintaining the order state, may be considered as plastic fashion.
In other world, the order state exists because of the
concentration of strength in the medium. In this case the symmetry
regulates the character of this concentration. At general in the
process of plastic deformation may exist as such situation :

Formation of order state = Spontaneous breaking symmetry =
Reconstruction of symmetry.

However, in this situation the reconstruction of symmetry turns to
no reduction of symmetry nut to the q-symmetry. Where we, in
secret sense, identity the process of reconstruction of symmetry
with the process of "quantization" of group Lie in the deformation
bundle . The parameters $q$ plays the role of the parameter of
regulation. Because of the ghost was considered as the
right-invariant basic in the bimodule, then he was one of types of
the plastic deformation keeping finite gauge-invariant character
of the functional $Q$.

\end{document}